\documentclass[superscriptaddress,twocolumn,showpacs,a4paper,
amssymb,amsmath,nobibnotes,aps,prd,showkeys,nofootinbib,notitlepage]{revtex4-1}

\usepackage{graphicx,subfigure,bm,color,psfrag}
\usepackage{amsfonts}
\usepackage{amsmath,latexsym}
\usepackage{mathtools}
\usepackage{verbatim}
\usepackage[arrowdel]{physics}
\usepackage{accents}
\usepackage{xspace}
\usepackage[pscoord]{eso-pic}
\usepackage[normalem]{ulem}
\usepackage[dvipsnames]{xcolor}
\usepackage[colorlinks = true, linkcolor = purple, urlcolor = purple, citecolor = blue, anchorcolor = blue]{hyperref}
\newcommand\e{\mathrm{e}}
\newcommand{\bea}{\begin{align}}
\newcommand{\eea}{\end{align}}

\newcommand\be{\begin{align}}
\newcommand\ee{\end{align}}
\begin{document}
\title{{Cosmology in $R^2$-gravity: Effects  of a Higher Derivative Scalar Condensate Background}}

\author{Raj Kumar Das}
\email{raj1996cool@gmail.com}
\affiliation{Physics and Applied Mathematics Unit, Indian Statistical Institute, 203 B. T. Road, Kolkata 700108, India}

\author{Aurindam Mondal}
\email{aurindammondal99@gmail.com}
\affiliation{Physics and Applied Mathematics Unit, Indian Statistical Institute, 203 B. T. Road, Kolkata 700108, India}

\author{Subir Ghosh}
\email{subirghosh20@gmail.com}
\affiliation{Physics and Applied Mathematics Unit, Indian Statistical Institute, 203 B. T. Road, Kolkata 700108, India}

\author{Supriya Pan}
\email{supriya.maths@presiuniv.ac.in}
\affiliation{Department of Mathematics, Presidency University, 86/1 College Street, Kolkata 700073, India}
\affiliation{Institute of Systems Science, Durban University of Technology, PO Box 1334, Durban 4000, Republic of South Africa}

\begin{abstract}
A well known extension of Einstein General Relativity is the addition of an $R^2$-term, which is free of ghost excitations and in the linearized framework, reduces Einstein General Relativity and an additional higher derivative scalar. According to \cite{Chakraborty:2020ktp}, the above scalar sector can sustain a Time Crystal-like
minimum energy state, with non-trivial time dependence.  Exploiting previous result that the scalar can sustain modes with periodic time dependence in its lowest energy, we consider this condensate as a source and study the Friedmann-Lema\^{i}tre-Robertson-Walker (FLRW) cosmology in this background. The effect of the $R^2$-term is interpreted as a back reaction. A remarkable consequence of the condensate is that, irrespective of open or close geometry of the Universe, for an appropriate choice of parameter window, the condensate can induce a decelerating phase before the accelerated expansion starts and again, in some cases, it can help to avoid the singularity in the deceleration parameter (that is present in conventional FLRW Cosmology). 
\end{abstract}

\maketitle
\section{Introduction}

Modern cosmology is full of excitements, thanks to a large amount of observational data from various astronomical missions. The physics of the early- and late universe has remained mysterious and this has created an enormous interest in the scientific community. Over the last several years, cosmologists are trying to understand the dynamics of our universe that should be consistent with the available observational probes at high and low redshifts. However, despite many observational missions, actual dynamics of the universe is yet to be discovered. The dynamics of the universe is hidden within the Einstein's gravitational equations which connect the geometry of the space-time with the matter distribution. In the background of a homogeneous and isotropic universe in its large scales~\cite{Wu:1998ad,Yadav:2005vv} characterized by the Friedmann-Lema\^{i}tre-Robertson-Walker (FLRW) geometry, one can investigate the dynamics of the universe provided the gravitational theory and the distribution of the matter are prescribed. In the context of dark energy (exotic fluid within the context of Einstein's General Relativity) \cite{Copeland:2006wr,Bamba:2012cp} or modified gravity theories ~\cite{Nojiri:2006ri,Capozziello:2007ec,Padmanabhan:2007xy,Sotiriou:2008rp,Silvestri:2009hh,DeFelice:2010aj,Nojiri:2010wj,Clifton:2011jh,Hinterbichler:2011tt,Capozziello:2011et,Sami:2013ssa,deRham:2014zqa,Cai:2015emx,Nojiri:2017ncd,Bahamonde:2021gfp} different dynamics of the universe have been proposed. While the dark energy and modified gravity models have been widely  confronted with the available observational data, but we still have several questions that are either weakly understood or unanswered. The initial singularity is one of them where the physical laws do not hold. This motivated the cosmologists to find an alternative theory that can avoid the initial singularity and becomes consistent with the observational evidences. The time periodic cosmology is one of such attempts.  
The time periodic evolution of cosmology, in the form of cyclic cosmology, introduced by  Steinhardt and Turok \cite{Steinhardt:2002ih}, created a lot of interest in the scientific community as it could address the issues of homogeneity, flatness, density fluctuations and most importantly could resolve the singularity problem in cosmology. Related previous works  were not satisfactory because as the scale factor $a(t)$  of the FLRW universe 
neared zero, matter and radiation densities became close to Planck density, (thus depending on unknown Quantum gravity effects), such that the bouncing property of $a(t)$ was adversely affected (for a discussion see eg.\cite{tolman1987relativity}). However, in \cite{Ijjas:2016tpn} the universe appears in a never-ending sequence of   `bang'
and  `crunch' phases such that  the temperature and density of the universe remain finite at each transition from crunch to bang. Interestingly enough, the model proposed in \cite{Ijjas:2016tpn} has a higher derivative kinetic terms along with a cubic Galileon field, which is relevant in our context since a higher derivative scalar (without the Galileon term) is present in our model as well. A further refined version of cyclic model for Cosmology has appeared in  \cite{Ijjas:2019pyf}.

In this perspective, we introduce a novel version of Cosmology, generated from a particular form of $f(R)$ gravity containing terms up to quadratic in $R$ ($R$ being the Ricci scalar) that has an inherent time periodic feature. Being a higher derivative theory {\it{without}} the ghost problem and supporting inflation, this type of models, proposed by Starobinsky \cite{Starobinsky:1980te}, generated a lot of activity, especially in the area of Cosmology. As is customary, the excitation spectrum was revealed in a linearized framework  \cite{Alvarez-Gaume:2015rwa}. Very interestingly and crucial for our purpose,  it was shown that, the linearized model in de Sitter (dS)  or anti de Sitter (AdS)  background,  decouples  into conventional spin-2 graviton and a {\it{higher derivative scalar (\mbox{HDS}) mode}}.  This additional scalar mode is already an embarrassment; on top of that it is higher derivative in nature. Of  course, any energy distribution will induce a gravitational interaction through its energy-momentum tensor. In the present work,  we would like to interpret the effect of the HDS on cosmological evolution, in a novel way, as a back reaction of the $R^2$-term on conventional Einstein gravity. Specifically, we analyse the  FLRW equations in presence of this HDS, with its energy-momentum tensor as  the  source. In an earlier study~\cite{Chakraborty:2020ktp},  (involving one of the present authors),  it was shown that the {\it{lowest energy state}} of this HDS, due to its non-canonical higher derivative structure, can sustain an explicitly time dependent (periodic in time) mode. It is worth emphasizing that no external and ad hoc source was introduced in  the model ~\cite{Alvarez-Gaume:2015rwa} considered here and the HDS is actually a combination of metric tensor field components; higher derivative nature of $R^2$-term is responsible for this extra degree of freedom. Hence, it is only natural to consider this mode as  a ripple in the spacetime fabric itself, thereby introducing a new dimensional scale - frequency of the HDS mode. Indeed, we expect to reveal qualitatively new phenomena since the lowest energy state of the  HDS source  is dynamical in time with an explicit expression for frequency as derived here. Since we have provided explicit expression for the frequency of the lowest energy mode in terms of $R^2$-gravity parameters, the corrections derived here will obviously depend on these parameters.

{Lastly we would like to speculate about a possible connection between the lowest energy mode of the HDS discussed above and Time Crystal  (TC) (for details see \cite{Shapere:2012nq, Wilczek:2012jt,Sacha:2017fqe}). As noted in \cite{Chakraborty:2020ktp}, the lowest energy mode of HDS was identified as a condensate having TC  characteristics. 
~In non-commutative extension of General Relativity \cite{Das:2018bzx} and in cosmological context this has created a fair amount of interest \cite{Bains:2015gpv,Bubuianu:2019gbo}. However, the present work is closer in spirit to an alternative formulation of Time Crystal-like system, developed by one of the present authors \cite{Ghosh:2012yj}, where models with higher time derivatives in a quadratic action can exhibit a Spontaneous Symmetry Breaking (SSB) in {\it{momentum space}} that induces a non-zero Fourier mode for the lowest energy state.   This happens due to the simultaneous presence of $|K|^4$ and $|K|^2$  terms in the energy and momentum space, $|K|$ being the momentum or energy. In fact, the phenomenon is quite similar to conventional SSB in coordinate space where quartic and quadratic potentials can generate a condensate with lower energy. Following this approach, it was demonstrated in \cite{Chakraborty:2020ktp} that HDS acts like a Time Crystal and precisely this system is being considered here in the cosmological context of  $R^2$-gravity.   }
 
{ Although this is not our major concern but regarding possible connection with TC mentioned above, in the context of Cosmology a clarification is required. As we are in purely classical regime we will refer to the state in question as the lowest energy solution. Indeed, it is well-known that defining energy in the context of Gravity or General relativity is problematic. There can be two ways to explain the minimum energy state in the present case: (i) note that using the linearized form of  $R^2$-gravity as in \cite{Alvarez-Gaume:2015rwa} we simply have a graviton plus HDS theory and we minimize the energy with the most reasonable restriction, i.e. positivity and (ii) in \cite{Bains:2015gpv,Easson:2016klq}, the external matter content is such that instead of settling down to a static minimum energy configuration, a Time Crystal like state is approached that  oscillates indefinitely. This is akin to our model with the important difference that HDS matter content is induced by the quadratic $R$ gravity internally (and not  introduced externally). In all these lowest energy states,  continuous translation symmetry is broken partially, due to the periodicity in time. }

{ In the present work we follow up the natural next step: direct effects of this Higher Derivative Scalar Condensate (HDSC), as a back reaction, in cosmology.} We compute the energy-momentum tensor for the HDSC and treat it as a source for Einstein's General Relativity (GR), with the $R^2$-term giving rise to a back reaction.  How will this TC condensate impact on the cosmological scenario? To answer this question we set up the Friedmann-Lema\^{i}tre-Robertson-Walker (FLRW) equations in presence of this HDSC source. Indeed, we expect and reveal qualitatively new phenomena since the HDSC is dynamical in time with an explicit expression for frequency as derived here.

 The article has been structured as follows. In section \ref{sec-2} we introduce the gravitational action for the quadratic gravity and the emergence of the HDSC. In section \ref{sec-3} we discuss the minimization of the 
energy-momentum tensor arising from the quadratic gravity and the HDSC states. In section \ref{sec-Friedmann} we present the Friedmann equations in a homogeneous and isotropic background for the current theoretical framework and present the analytical solutions for the cosmological parameters. Then in section \ref{sec-results} we describe the cosmological dynamics in the HDS background in terms of the evolution of the key cosmological parameters and the implications of the results. Finally, in section \ref{sec-conclusion}, we close the present article with a brief summary of the entire work, major conclusions and future directions.  

\section{$R^2$ Gravity and Higher Derivative Scalar Condensate}
\label{sec-2}

Before proceeding towards the $R^2$ model, let us consider a generic form of  higher derivative action for a scalar field $\phi$ in curved background  (see for example \cite{Gibbons:2019lmj})
\begin{equation}
S=-\frac{1}{2}\int d^4x~\sqrt{-g} \bigg[(\Box\phi)^2+(m_1^2+m_2^2)g^{\mu\nu}\partial_\mu \phi\partial_\nu \phi +m_1^2m_2^2\phi^2 \bigg]
\label{1}
\end{equation}
for which the equation of motion and energy-momentum tensor are respectively
\begin{equation}
 \big[(\Box -m_1^2)(\Box -m_2^2)\big]\phi =0,
\label{3}
\end{equation}
\begin{eqnarray}
T_{\mu\nu} = -(\partial_\mu\Box\phi)\partial_\nu\phi -(\partial_\nu\Box\phi)\partial_\mu\phi \nonumber \\ +  \frac{g_{\mu\nu} (\Box\phi)^2}{2} -\frac{m_1^2m_2^2}{2} g_{\mu\nu}\phi^2  +g_{\mu\nu}g^{\alpha\beta}(\partial_\alpha\Box\phi )\partial_\beta\phi  \nonumber \\ +(m_1^2+m_2^2) \bigg(\partial_\mu\phi\partial_\nu\phi -\frac{1}{2}g_{\mu\nu}g^{\alpha\beta}\partial_\alpha \phi\partial_\beta\phi \bigg)
\label{10aa}
\end{eqnarray}
This expression will play an essential role in the subsequent analyses.
The QG action we focus on  is given by \cite{Alvarez-Gaume:2015rwa}
\begin{equation}
A=\frac{c^4}{16\pi G}\int d^4x~ \sqrt{-g}\bigg[R+ \alpha R^2-2\Lambda \bigg]
\label{4}
\end{equation}
In order to avoid tachyonic excitations, conventionally $\alpha$ is taken to be positive. In Ref. \cite{Alvarez-Gaume:2015rwa}, the authors have shown that this $R^2$ action decouples to  conventional gravity theory (with a spin $2$ graviton) along with a higher derivative scalar sector (modulo surface terms). Very interestingly, higher derivative nature of the initial $R^2$ action becomes confined to the decoupled scalar sector given below
\begin{eqnarray}
\frac{A_{HDS}}{\Sigma}= -\frac{1}{2}\int d^4x~\sqrt{-\widetilde g}\Bigg[(\widetilde{\Box} \phi )^2 -\frac{ R_{0}}{18\alpha}\phi ^2 \nonumber \\ + \left(\frac{ R_{0}}{3}-\frac{1}{6 \alpha} \right)\phi\widetilde{\Box}\phi \Bigg]
\label{8aa}
\end{eqnarray}
where $\Sigma =(-\frac{9\alpha c^4}{8.16\pi G})$, $\widetilde {g}_{\mu\nu}$ is the arbitrary background metric (although in our case we restrict to dS or AdS), $\widetilde{\Box}$ is defined accordingly. We consider $ R_{0}$ to be the constant curvature corresponding to the maximally symmetric background spacetime $\widetilde {g}_{\mu\nu}$ \cite{Chakraborty:2020ktp,Alvarez-Gaume:2015rwa}. 

Exploiting (\ref{10aa}) the energy-momentum tensor for (\ref{8aa}) becomes 
\begin{eqnarray}
\frac{T_{\mu\nu}}{\Sigma} = -(\partial_\mu\Box\phi)\partial_\nu\phi -(\partial_\nu\Box\phi)\partial_\mu\phi + \frac{1}{2}g_{\mu\nu}(\Box\phi)^2 \nonumber \\ +g_{\mu\nu}g^{\alpha\beta}(\partial_\alpha\Box\phi )\partial_\beta\phi 
+ \left(\frac{1}{2} \right) \frac{ R_{0}}{18 \alpha}g_{\mu\nu}\phi^2 \nonumber\\ + \left(\frac{1}{6\alpha} -\frac{ R_{0}}{3} \right) \left(\partial_\mu\phi\partial_\nu\phi -\frac{1}{2}g_{\mu\nu}g^{\alpha\beta}\partial_\alpha \phi\partial_\beta\phi \right) 
\label{10a}
\end{eqnarray}
where we identify the parameters  as 
 \begin{equation}
     (m_1^2+m_2^2)=\bigg(\frac{1}{6\alpha}-\frac{ R_{0}}{3} \bigg) \hspace{0.2cm}; \quad 
      m_1^2m_2^2=\bigg(-\frac{ R_{0}}{18\alpha} \bigg)
      \label{par}
 \end{equation}
It is straightforward to check the on-shell conservation law $\nabla _\mu T^{\mu\nu}=0$.
We consider the  unit system used, $l~(\mbox{length})$, $m~(\mbox{mass})$, $t~(\mbox{time})$ such that dimension of $[T_{\mu\nu}] = (ml^2/t^2) (1/l^3) = \mbox{energy~density}.$
 For real values for $m_1,m_2$, and for  $\alpha>0$, we impose $ R_{0}= -|R_{0}|$ ($\sim$ an AdS background), and  obtain
 \begin{equation}
     m_1=\sqrt{\frac{1}{6\alpha}} \hspace{0.34cm} ; \quad 
      m_2=\sqrt{\frac{|R_{0}|}{3}}
      \label{mm}
 \end{equation}
 Since $\phi$ is decoupled from the Einstein gravity sector, we propose to interpret $T_{\mu\nu} (\phi)$ as a source, and using the Einstein's gravitational equations, cosmology in presence of $T_{\mu\nu} (\phi)$ can be explored. The main novelty of our scheme is that this ``matter'' sector is not introduced from outside but it is actually generated by the $\alpha R^2$ term in the $R^2$ action. Thus,  the  Einstein's equations for this set-up 
 read
\begin{equation}
\bigg(R_{\mu\nu}-\frac{1}{2}g_{\mu\nu}R \bigg)=\frac{8\pi G}{c^4} T_{\mu\nu}
\label{6}
\end{equation}
where the left hand side is provided by the Einstein's GR action $\sim \int \sqrt{-g}R$, $T_{\mu\nu} (\phi)$ in the right hand side,  will be given explicitly  by using the specific  HDSC solution for $\phi$
in the next section.

 \section{Minimization of the energy-momentum tensor and TC ground states}
 \label{sec-3}

  We quickly recapitulate the previous work \cite{Chakraborty:2020ktp} (involving one of the present authors) to find the lowest energy HDSC  solutions. In particular, this means that {\it{the  vacuum is replaced by the TC condensate with the latter acting as a stage for Einstein gravity}}. Explicit form of the HDSC parameters will be obtained by minimizing the energy $\widetilde{T}_{00}$ in momentum space. Let us consider a superposition of plane waves of frequency $\omega$ and wave number $K$,
  \begin{equation}
  \label{ph}
    \widetilde\phi(K)=\int \frac{d^4x}{(2\pi)^4}\e^{-i(\omega\eta-K.x)}\phi(x) 
 \end{equation}
Using the above, $T_{00}$ from eqn. (\ref{10a}) in the $K$-space is given by
 \begin{equation}
    \frac{\widetilde T_{00}}{\Sigma}= f(\omega,K)|\widetilde\phi(K)|^2 
 \end{equation}
 where $f(\omega,K)$ takes the following form
 
 \begin{eqnarray}
     f(\omega,K)= -\alpha\Bigg[\frac{(-3\omega^4+2c^2 K^2\omega^2+c^4K^4)}{2c^2a^2} \nonumber\\+\frac{(m_1^2+m_2^2)(\omega^2+c^2K^2)}{2}+\frac{c^2a^2(m_1m_2)^2}{2}\Bigg]
     \label{ab}
 \end{eqnarray}
In the following we impose two dispersion relations. 
\begin{itemize}
    \item  {\it Dispersion relation  I:}  From eqns. (\ref{3}) and (\ref{mm}), the first dispersion relation is given by
 \begin{equation} 
    \omega^2= c^2\bigg(K^2+\frac{|R_{0}|a^2}{3}\bigg).
 \end{equation}
 Using this in (\ref{ab}) we minimise $\widetilde T_{00}(K)$ from $\frac{d}{dK}\left(\frac{\widetilde T_{00}}{\Sigma} \right)=0; \quad
     \frac{d^2}{dK^2} \left(\frac{\widetilde T_{00}}{\Sigma} \right)>0$ (or $\widetilde T_{00}(\omega)$ for $\omega$) to find  the following TC parameters
 \begin{equation}
 \label{sol}
     K_0=0 \hspace{0.24cm} ; \quad \omega_0^2=\bigg(\frac{c^2|R_{0}|a^2}{3} \bigg)
 \end{equation}
subject to the constraints $\alpha>0, |R_{0}|>0,  (1-2\alpha |R_{0}|)<0$. 
 Substituting the above results,  minimum energy density for the condensate  appears as
 \begin{eqnarray}
\frac{\widetilde T_{00}|_{min}}{\Sigma} = -\alpha\Bigg[K_0^2 \left(\frac{1}{6\alpha}-\frac{|R_{0}|}{3} \right) \nonumber \\ +\frac{|R_{0}|a^2}{6} \left(\frac{1}{3\alpha}-\frac{2|R_{0}|}{3} \right)\Bigg]|\widetilde\phi|^2 \nonumber \\
= -\frac{|R_{0}|a^2}{18} \left(1-2\alpha |R_{0}| \right)|\widetilde\phi|^2
 \end{eqnarray}
 which  is positive since  $(1-2\alpha |R_{0}|)$ is negative. Thus, we reveal the important result that in quadratic gravity, the $R$ and $R^2$ terms can conspire to generate a {\it{stable lowest energy condensate state}} $-$ a HDSC.

 \item {\it Dispersion relation  II:}  In a similar way,  dispersion relation II yields  the minimization condition  $ K_0=0,  \omega_0^2=\frac{c^2a^2}{6\alpha}$ 
 with  $\alpha>0, |R_{0}|>0, (1-2\alpha |R_{0}|)>0$.
 Once again we recover a  minimum positive energy density condensate 
  \begin{eqnarray}
     \frac{\widetilde T_{00}|_{min}}{\Sigma} &=&\alpha \omega^2 \left(\frac{1}{6\alpha}-\frac{|R_{0}|}{3} \right)|\widetilde\phi|^2 \nonumber \\
      &=& \frac{a^2}{36\alpha} \left(1-2\alpha |R_{0}| \right)|\widetilde\phi|^2 .
 \end{eqnarray}

\end{itemize}
 
\section{Friedmann equations in HDSC background} 
\label{sec-Friedmann}

The Friedmann–Lema\^{i}tre–Robertson–Walker(FLRW) metric in 
conformal time $\eta$ is given by 
\begin{equation}
\label{fc}
    ds^2= a^2(\eta)\bigg[-c^2d\eta^2+\frac{dr^2}{1-kr^2}+r^2(d\theta^2+\sin^2\theta d\phi^2)\bigg], 
\end{equation}
where $a(\eta)$ is the expansion scale factor of the universe  and $k$ denotes the spatial geometry of the universe;  $k =0,+1,-1$, represents a flat, closed and open  universe, respectively. Einstein's  equations (\ref{6}) reproduces the Friedmann equations
\begin{eqnarray} 
 \left(\frac{a'}{a} \right)^{2}=\bigg[-\frac{8\pi G a^2}{3}T^0_0+\frac{c^2\Lambda}{3}a^2-kc^2 \bigg], \label{a4} \\
  \frac{a''}{a}- \left(\frac{a'}{a} \right)^2= \bigg[\frac{4\pi Ga^2}{3}(T^0_0-3T^i_i)+\frac{a^2c^2\Lambda}{3} \bigg], \label{a44}
\end{eqnarray}
where $a' \equiv da/d\eta $. Components of  the energy-momentum tensor appearing in eqns. (\ref{a4}, \ref{a44}) are 
\begin{align}
  \widetilde{T}^0_0=\Sigma\Bigg[\frac{1}{c^6a^8} \left(\frac{\phi''^2}{2}+4\phi'\phi''\frac{a'}{a}+\frac{11\phi'^2}{2}\frac{a'^2}{a^2}-\phi'^2\frac{a''}{a}-\phi'\phi''' \right) \nonumber\\-\left(\frac{m_1^2+m_2^2}{2c^4a^4} \right)\phi'^2- \left(\frac{m_1^2m_2^2}{2c^2} \right)\phi^2\Bigg],
\end{align}

\begin{align}
  \widetilde{T}^i_i=\Sigma\Bigg[\frac{1}{c^6a^8}\left(\frac{\phi''^2}{2}-2\phi'\phi''\frac{a'}{a}-\frac{9\phi'^2}{2}\frac{a'^2}{a^2}+\phi'^2\frac{a''}{a}+\phi'\phi''' \right) \nonumber\\+ \left(\frac{m_1^2+m_2^2}{2c^4a^4} \right)\phi'^2- \left(\frac{m_1^2m_2^2}{2c^2} \right)\phi^2\Bigg],
\end{align}
where $\phi' \equiv d\phi/d\eta$.  We derive a quadratic equation from eqns. (\ref{a4}), (\ref{a44})  
\begin{eqnarray}
      \left(\frac{a'}{a} \right)^{2}-\frac{3\alpha\phi'\phi''}{4c^2a^6} \left(\frac{a'}{a} \right)- \Gamma = 0,
\end{eqnarray}
where $ \Gamma $ is given by
\begin{equation}
    \Gamma = \bigg(\frac{c^2\Lambda a^2}{3}-\frac{\phi'^2}{64a^2}+\frac{c^2a^2R_{0}\phi^2}{192}-kc^2\bigg) G(\alpha),
    \end{equation} 
    \begin{widetext}
    in which 
 \begin{equation}
          G(\alpha)= \Bigg[1- \frac{3\alpha}{16c^2a^6}\; \left( \frac{\frac{13\phi'^4}{128}+\frac{\phi'^2}{2}\bigg(9kc^2-R_{0}c^2a^4-\frac{7c^2\Lambda a^2}{3} \bigg)-\frac{7}{2} \bigg(\frac{c^2a^2R_{0}\phi^2\phi'^2}{192} \bigg)-\bigg(\frac{\phi''^{2}-2\phi'\phi'''}{2} \bigg)}{\bigg(\frac{c^2\Lambda a^2}{3}-\frac{\phi'^2}{64a^2}+\frac{c^2a^2R_{0}\phi^2}{192}-kc^2 \bigg)} \right) \Bigg].
\end{equation}
\end{widetext}
Note that in absence of $\alpha R^2$ term, the HDSC $\phi$ is not generated and we recover the Friedmann equations with curvature and the cosmological constant.

In order to proceed further we consider some approximations: 
(I) we assume small $\alpha$ so that the effects of $\alpha R^2$ term can be treated perturbatively, i.e. $O(\alpha^2) \approx 0$ and we get 
\begin{equation}
    \frac{a'}{a} = \bigg(\frac{3\alpha}{8c^2a^6}\phi'\phi''\pm \sqrt{\Gamma} \bigg) .
    \label{r1}
\end{equation}
Note that due to the small $\alpha$ approximation, it will not be appropriate to use $m_2$ in (\ref{mm}) and we will use $m_1$ only. (II)  secondly, putting back eq. (\ref{sol}) in eq. (\ref{ph}) and considering the real part yields  $\phi \sim \cos(\omega\eta)$, where $\omega = ca\sqrt{\frac{|R|}{3}}$. We use this explicit form of $\phi$ to compute  $a'/a$ in (\ref{r1}) and furthermore,   make a  small $\eta$  approximation such that  $\sin(\omega\eta) \sim 0 ;~ \cos(\omega\eta) \sim 1$ to    recover a  tractable result for the Hubble parameter in conformal time.  

For subsequent work we switch over to cosmic time $t$
and derive  the Hubble parameter
\begin{eqnarray}
     H=\frac{\dot a}{a} \boldsymbol{\approx} \pm \sqrt{\bigg(\frac{c^2\Lambda}{3}+\frac{c^2R_{0}}{192}-\frac{kc^2}{a^2}\bigg)}  \Bigg[1 + \nonumber \\   \left(\frac{\alpha R_{0}^2c^2}{192a^4} \right)  \left(\frac{1}{\frac{c^2\Lambda}{3}+\frac{c^2R_{0}}{192}-\frac{kc^2}{a^2}} \right)\Bigg] 
\end{eqnarray}
to first order in the parameter  $\alpha$, denoting the coupling of $R^2$ term.
With effective cosmological constant $\Lambda_{\rm eff} = \Lambda + \frac{|R_{0}|}{64}$, the Hubble equation is rewritten as 
\begin{equation}
\label{hh}
     H^2=\frac{\dot a^2}{a^2} \approx  \bigg(\frac{c^2\Lambda_{\rm eff}}{3}-\frac{kc^2}{a^2}+ \alpha\frac{R_{0}^2c^2}{96a^{4}} \bigg).
\end{equation}  
An interesting observation is that, in our crude approximation, the TC condensate correction behaves like a radiation contribution, generated solely from metric. 

In a convenient parametrization, let us write
\begin{equation}
\label{hubble-explicit}
\frac{H(a)}{H_0}=\sqrt{\Omega_{\Lambda}+\Omega_{k} \left(\frac{a}{a_0} \right)^{-2}+\Omega_{\alpha} \left(\frac{a}{a_0} \right)^{-4}}
\end{equation}
where  eqn. (\ref{hh}) is used at $a= a_0$ (equivalently, $t=t_0$) to define
\begin{equation}
H_0^2=\bigg(\frac{c^2\Lambda_{eff}}{3}-\frac{kc^2}{a_0^2}+\frac{\alpha c^2|R_{0}|^2}{96a_0^4} \bigg).
\end{equation}
Here,  $\Omega_{\Lambda},\; \Omega_{k},\; \Omega_{\alpha}$ are the dimensionless density parameters of the associated fluid components at present time, defined as  
\begin{equation}
\label{om}
\Omega_{\Lambda}=\frac{c^2\Lambda_{\rm eff}}{3H_{0}^2}, ~~
\Omega_{k}=-\frac{kc^2}{H_{0}^2a_{0}^2 }, ~~
\Omega_{\alpha}=\frac{\alpha c^2|R_{0}|^2}{96H_{0}^2a_{0}^2},
\end{equation}
From eq. (\ref{hubble-explicit}), we get the constraint relation $\Omega_{\Lambda}+\Omega_{k}+\Omega_{\alpha}=1$. 
Furthermore, an analytic solution of  scale factor $a(t)$ can be expressed in terms of  density parameters as
\begin{eqnarray}
a(t)=\Bigg[\frac{\chi}{2}e^{2H_{0}\sqrt{\Omega_{\Lambda}}(t-t_{\star})} + \left(\frac{\Omega_{k}^2}{8\Omega_{\Lambda}^2} -\frac{\Omega_{\alpha}}{2\Omega_{\Lambda}} \right)e^{-2H_{0}\sqrt{\Omega_{\Lambda}}(t-t_{\star})} \nonumber \\ -\frac{\Omega_{k}}{2\Omega_{\Lambda}} \Bigg]^{1/2},
\end{eqnarray}
where 
\begin{equation}
\chi(a_{\star})= \left(a_{\star}^2+\frac{\Omega_{k}}{2\Omega_{\Lambda}}\right)+\sqrt{\left(a_{\star}^2+\frac{\Omega_{k}}{2\Omega_{\Lambda}} \right)^2+\left(\frac{\Omega_{\alpha}}{\Omega_{\Lambda}}-\frac{\Omega_{k}^2}{4\Omega_{\Lambda}^2}\right)}.
\end{equation}
The other independent solution for $a(t)$ is
\begin{align}
a(t)=\Bigg(-\frac{\chi}{2}e^{2H_{0}\sqrt{\Omega_{\Lambda}}(t-t_{\star})} +\left(\frac{\Omega_{\alpha}}{2\Omega_{\Lambda}} -\frac{\Omega_{k}^2}{8\Omega_{\Lambda}^2} \right)e^{-2H_{0}\sqrt{\Omega_{\Lambda}}(t-t_{\star})} \nonumber\\-\frac{\Omega_{k}}{2\Omega_{\Lambda}}\Bigg)^{1/2}.
\end{align}

The deceleration parameter, defined in terms of  cosmic time  $q \equiv -1-\dot{H}/H^2 = -(a\ddot a)/(\dot a)^2$,
turns out to be
\begin{equation}
\label{q}
  q(a)=\bigg[-1+\left(\frac{\Omega_k(\frac{a}{a_0})^{-2}+2\Omega_\alpha(\frac{a}{a_0})^{-4}}{\Omega_\Lambda+\Omega_k(\frac{a}{a_0})^{-2}+\Omega_\alpha(\frac{a}{a_0})^{-4}}\right)\bigg]
\end{equation}
\begin{figure}[ht!]
    \centering
    \includegraphics[width=0.45\textwidth]{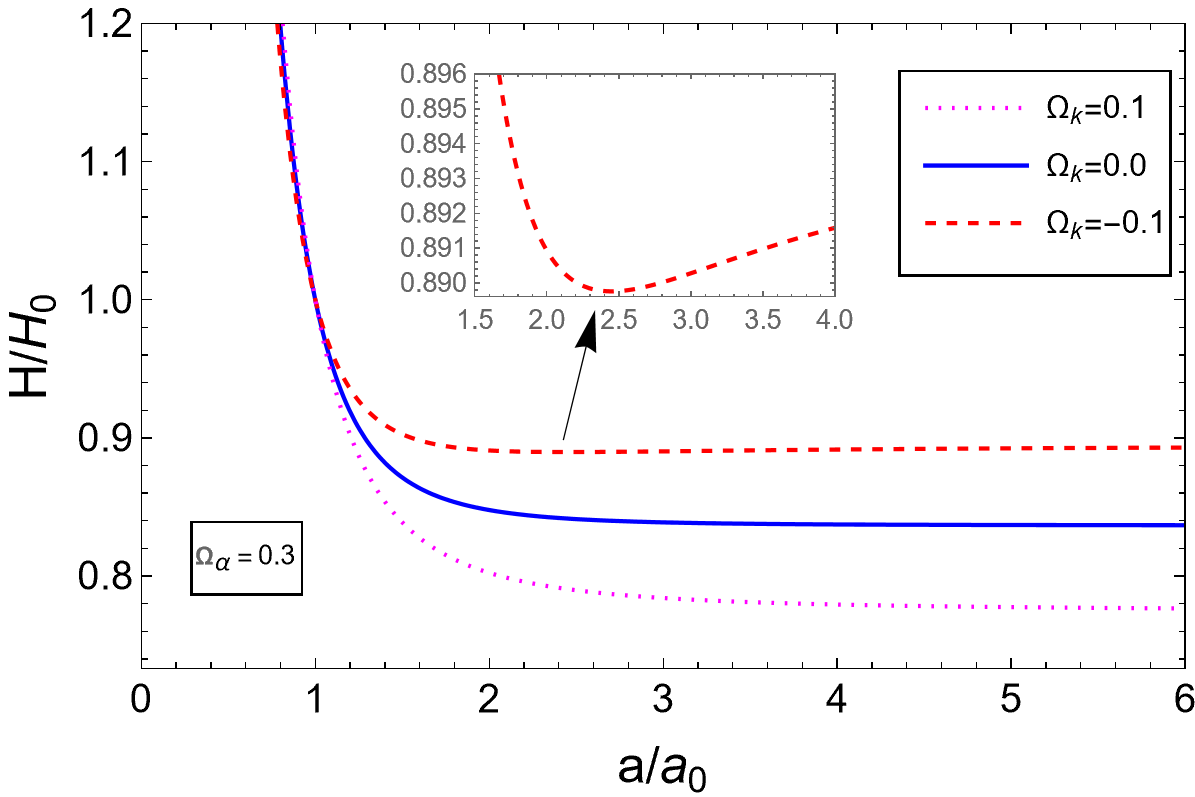}
    \caption{The evolution of the dimensionless Hubble function with respect to the cosmic scale factor for a fixed value of $\Omega_\alpha~(= 0.3)$ has been depicted for three spatial geometries of the universe, namely, open universe (magenta dotted curve), flat universe (blue solid curve),  and closed universe (red dashed curve). Here, $\Omega_{\Lambda} + \Omega_{k} +\Omega_{\alpha} =1$. }
    \label{Figure1}
\end{figure}
\begin{figure}
    \includegraphics[width=0.45\textwidth]{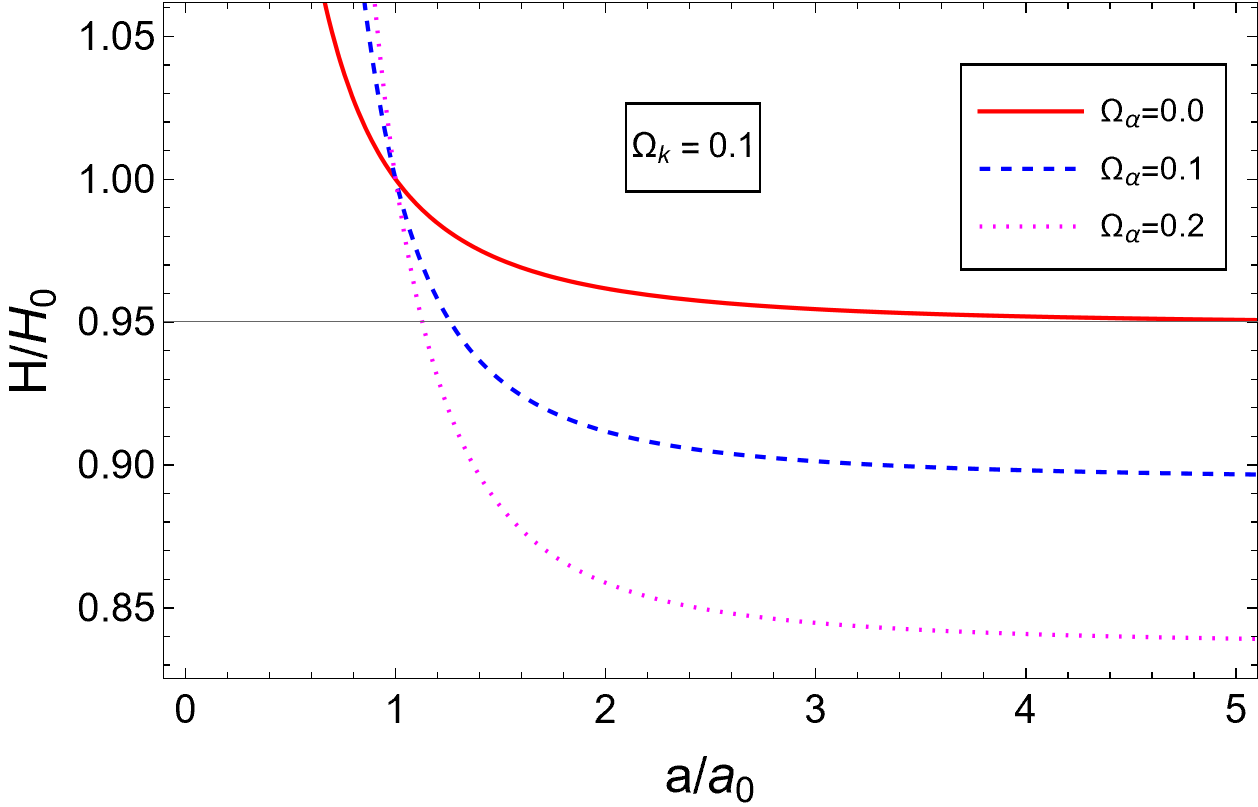}
    \includegraphics[width=0.45\textwidth]{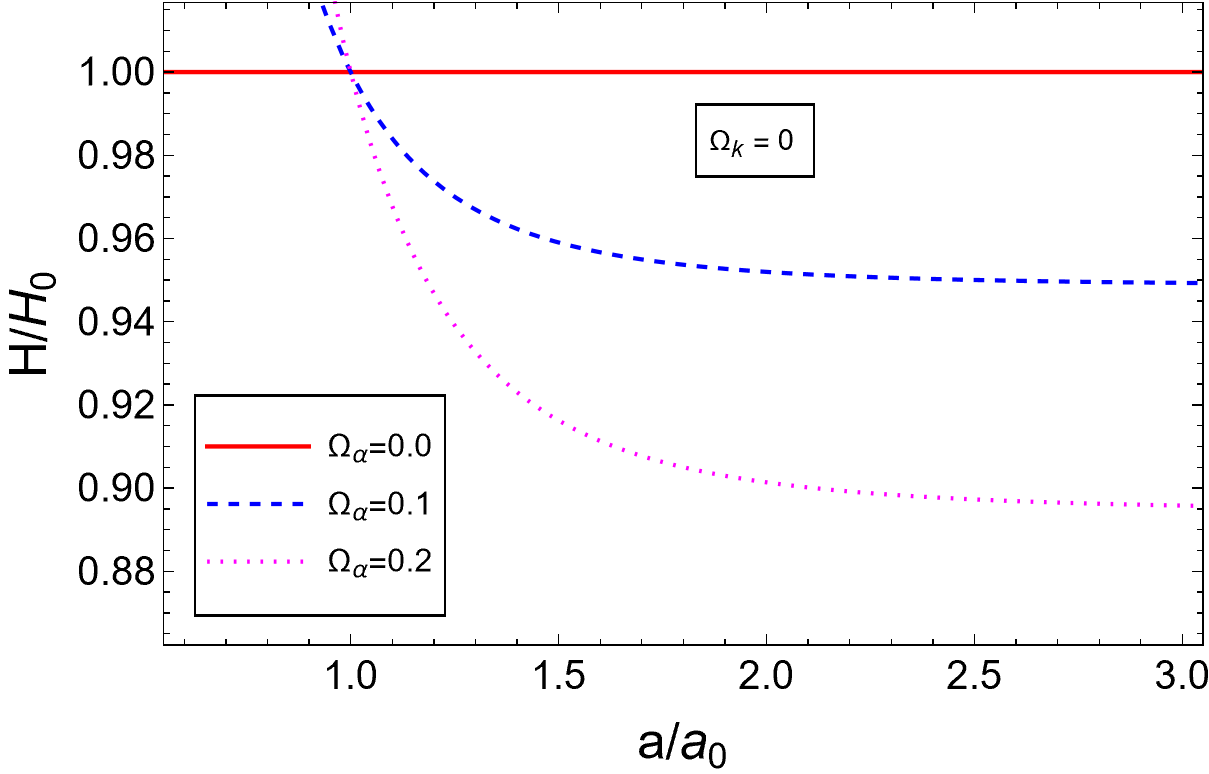}
    \includegraphics[width=0.45\textwidth]{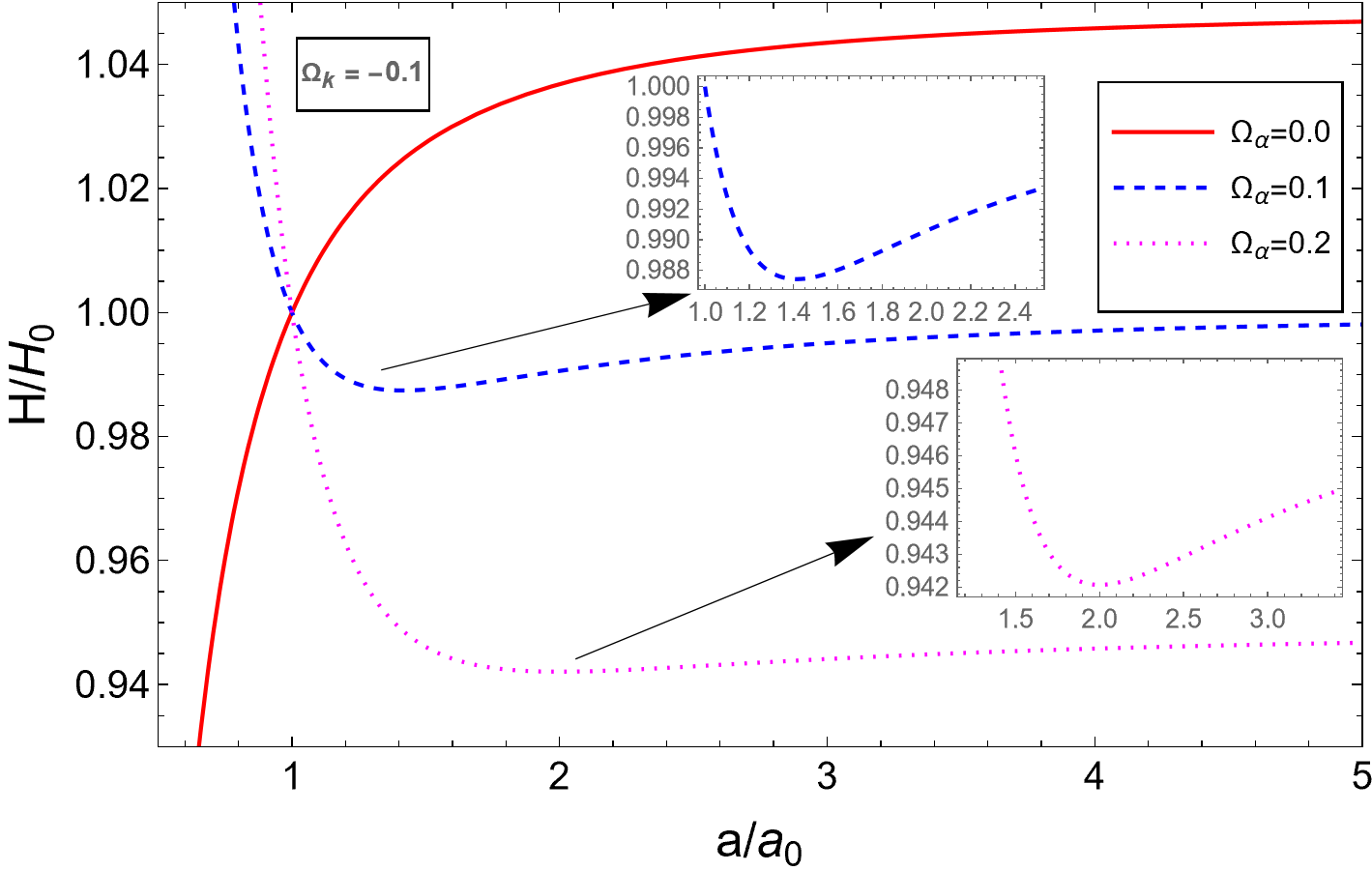}
    \caption{We display the evolution of the dimensionless Hubble function with the cosmic scale factor for different values of $\Omega_{\alpha}$ considering three spatial geometries of the universe, namely, open universe (upper graph), flat universe (middle  graph) and closed universe (lower graph). In each graph, the solid red curve corresponds to $\Omega_{\alpha} = 0$; dashed blue curve corresponds to $\Omega_{\alpha} =0$, and dotted magenta curve corresponds to $\Omega_{\alpha} =0.2$. Here, $\Omega_{\Lambda} + \Omega_{k} +\Omega_{\alpha} =1$. }
    \label{Figure234}
\end{figure}
\begin{figure}[!h]
    \centering
    \includegraphics[width=0.45\textwidth]{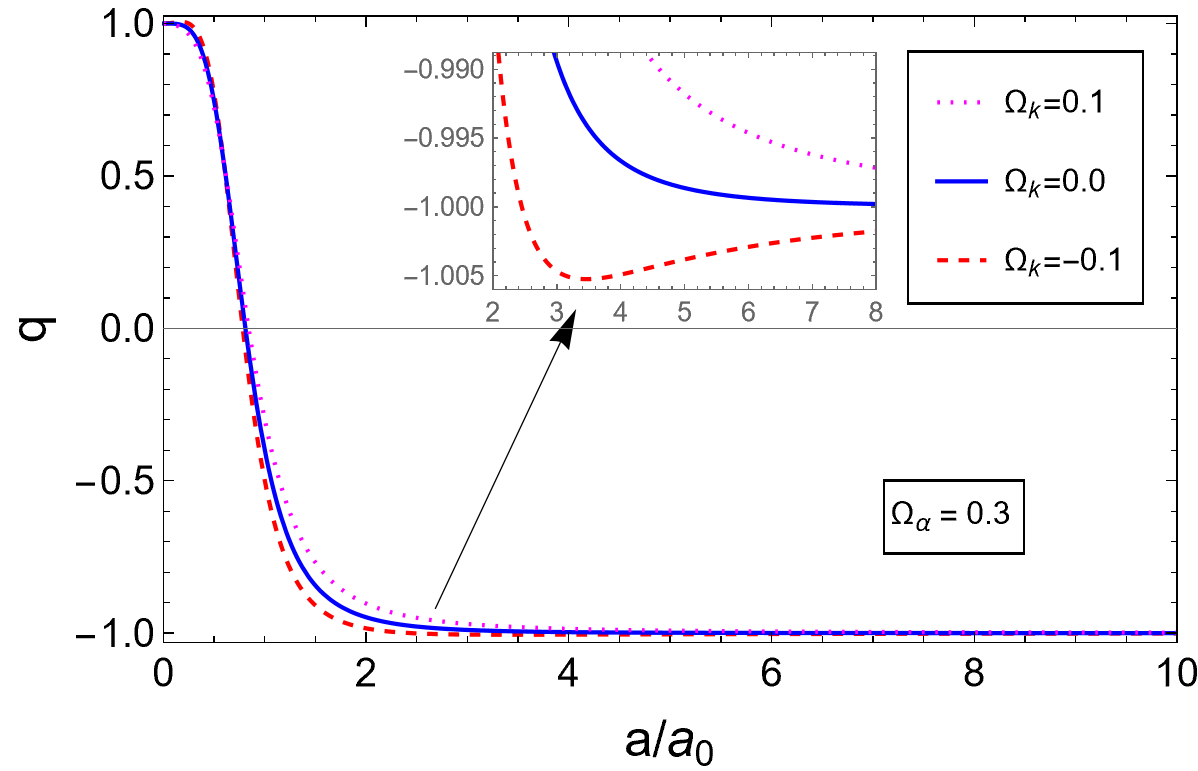}
    \caption{The evolution of the deceleration  parameter for a fixed value of $\Omega_\alpha~(=0.3)$ has been shown for different spatial geometries of the universe, namely, open universe (magenta dotted curve), flat universe (blue solid curve),  and closed universe (red dashed curve). Here, $\Omega_{\Lambda} + \Omega_{k} + \Omega_{\alpha} =1$. }
    \label{Figure5}
\end{figure}
\begin{figure}
    \centering
    \includegraphics[width=0.45\textwidth]{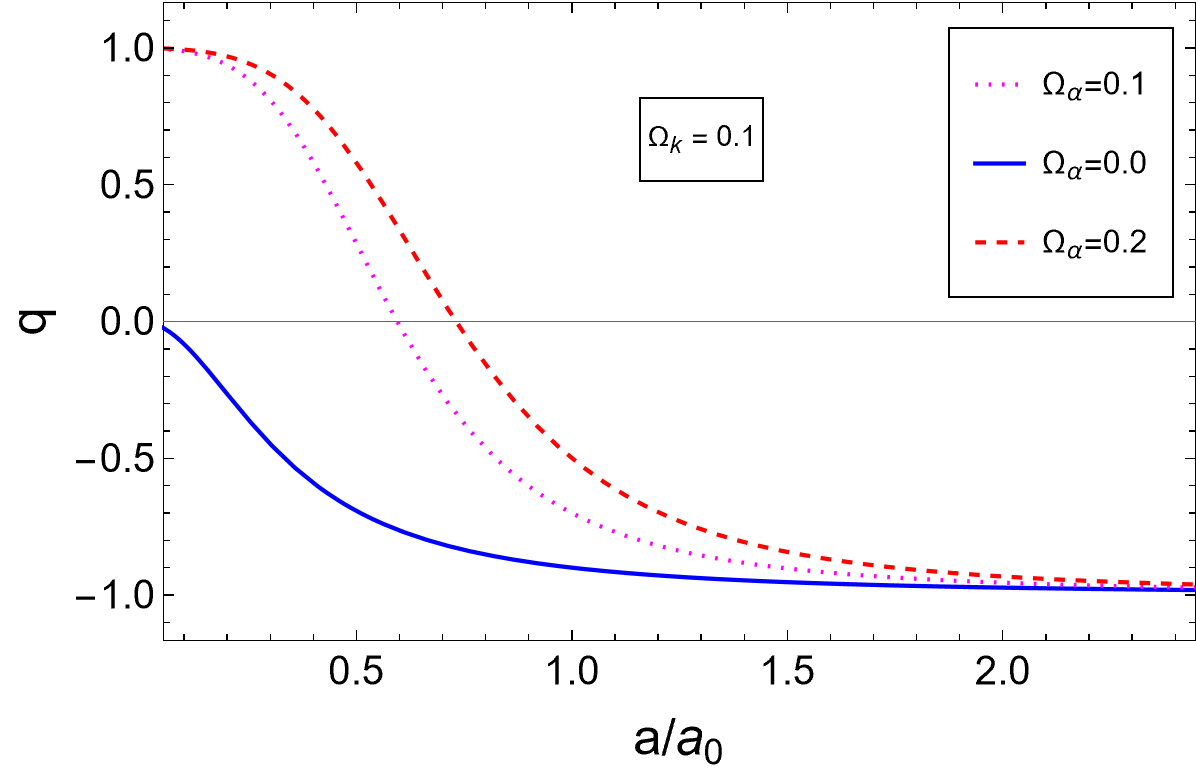}
    \includegraphics[width=0.45\textwidth]{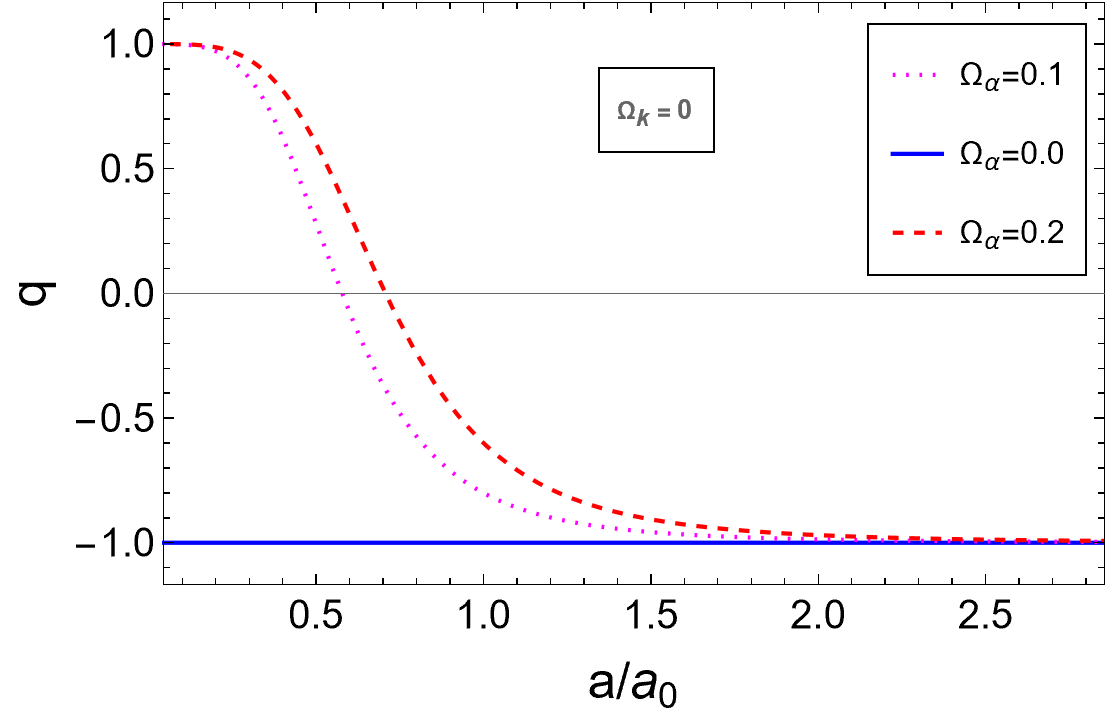}
    \includegraphics[width=0.45\textwidth]{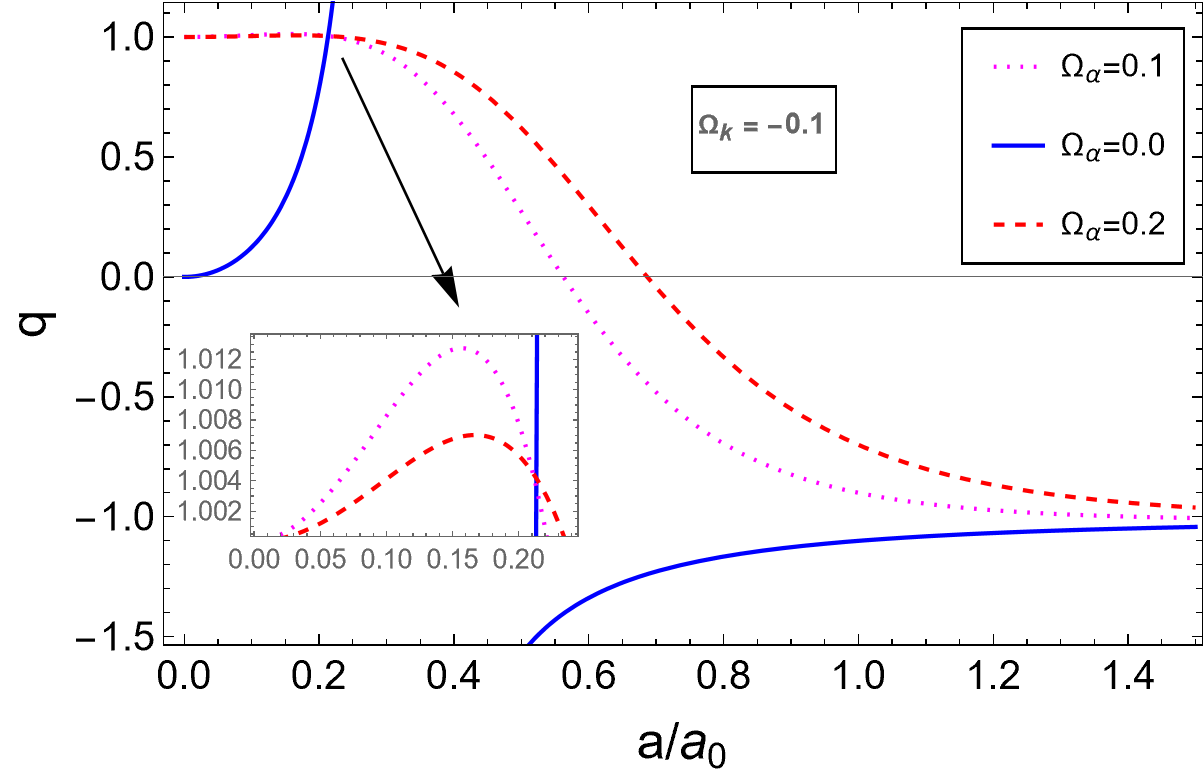}
    \caption{We present the evolution of the deceleration  parameter for different $\Omega_\alpha$ considering three spatial geometries of the universe, namely, open universe (upper graph), flat universe (middle graph) and closed universe (lower graph). In each graph, the solid red curve corresponds to $\Omega_{\alpha} = 0$, dashed blue curve corresponds to $\Omega_{\alpha} =0$ and dotted magenta curve corresponds to $\Omega_{\alpha} =0.2$. Here, $\Omega_{\Lambda} + \Omega_{k} + \Omega_{\alpha} =1$.  }
    \label{Figure678}
\end{figure}
\section{Results and their Implications}
\label{sec-results}

Let us finally discuss the major consequences  the novel Higher Derivative Scalar (or Time Crystal-like) condensate can have on the evolution of the universe. As described in section \ref{sec-Friedmann}, the HDSC background can provide an expression for the scale factor $a(t)$ in terms of density parameters $\Omega_j$ ($j = \Lambda, \alpha, k$). This leads to two of the well-known cosmological parameters, namely, the Hubble parameter $H(a)$ and the deceleration parameter $q(a)$ which offer a clear picture of the dynamics of our universe. In order to understand the dynamics of the universe, in Figs. \ref{Figure1}, \ref{Figure234}, \ref{Figure5}, \ref{Figure678}, we provide the graphical descriptions of $H(a)$ and $q(a)$ in various curvature scenarios and strengths of the HDSC through $\alpha$ (generated by the $\alpha R^2$-term in gravity). 
In all the figures, $\Omega_k$ can take positive, negative or zero value corresponding to open (negative curvature), closed (positive curvature) or flat (zero curvature) universe, respectively. And we note that for drawing the curves the values of the density parameters have been chosen in such a way so that the closure equation $\sum_{i} \Omega_i = 1$ is satisfied.

In Fig. \ref{Figure1}, for a fixed value of $\Omega_\alpha =0.3 $, the evolution of the dimensionless Hubble parameter $H/H_0$ is displayed against $a/a_0$ for positive, negative and vanishing $\Omega_k$. Since from eqn. (\ref{om}), it is clear that $\Omega_\Lambda, \Omega_\alpha$ are always positive but $\Omega_k$ can change sign as $k$ can take three distinct values ($0, +1, -1$), and only for the negative $\Omega_k$ (i.e. positive curvature or closed universe), it contributes oppositely and there is the possibility of a stationary point (in the present case a minimum) in $H(a)~ vs.~ a$ profile.

To establish the effects of HDSC clearly, in Fig.  \ref{Figure234}, we plot the evolution of the Hubble parameter in terms of $a/a_0$  for three distinct spatial geometries of the universe, namely,
$\Omega_k= 0.1$  (upper left graph of Fig. \ref{Figure234}), $\Omega_k=0$ (upper right graph of Fig. \ref{Figure234}) and $\Omega_k=-0.1$  (lower graph of Fig. \ref{Figure234}), respectively, and for each case we take three distinct values of $\Omega_{\alpha}$ as $\Omega_\alpha =0, 0.1,0.2$ keeping in mind that  it is positive and small and fixing $\Omega_\Lambda$ accordingly. Notice that  for $\Omega_k>0$ (open universe) the profiles are qualitatively similar (see the upper left graph of Fig. \ref{Figure234}) with only quantitative differences as $\Omega_\alpha$ increases. However, the graphs change significantly for $\Omega_k=0$ (flat universe) where  $\Omega_\alpha =0$ yields the well-known constant value of $H(a)$ but non-zero $\Omega_\alpha $'s show distinct variation of $H(a)$ with respect to $a$ (see the upper right graph of Fig. \ref{Figure234}). Once again, the $\Omega_k<0$, see the lower graph of Fig. \ref{Figure234} (closed universe) is the most interesting scenario where {\it{ non-zero values of $\Omega_\alpha$ generate completely different profiles compared to $\Omega_\alpha = 0$ (no HDSC);  the former ones have a minimum before saturating for large $a$}}.
It is also easy to see from eqn. (\ref{hubble-explicit}) that all the  curves for different $\Omega_\alpha $ will cross  at  $a=a_0$.

Now we study the behavior of the deceleration parameter $q(a)$ for the same sets of parameters as considered earlier. Fig.  \ref{Figure5} demonstrates a very interesting fact that for any universe (with $\Omega_k$ being positive, negative or zero), a non-trivial $\Omega_\alpha$ induces a {\it{change in the sign of}} $q(a)$ {\it{from positive to negative as}} `$a$' {\it{increases from 0}}. This indicates that the HDSC generates a decelerating phase before the acceleration starts. Clearly this indicates that after the decelerating phase the universe enters into the late time accelerating phase. 

Once again we analyse the strength of $\Omega_\alpha$ for the three different spatial structures of the universe. In Fig. \ref{Figure678} we have summarized the evolution of the deceleration parameter  for $\Omega_k$ being positive (corresponds to the upper left graph of Fig. \ref{Figure678}), zero (upper right graph of Fig. \ref{Figure678}) and negative (lower graph of Fig. \ref{Figure678}) respectively, taking three distinct values of $\Omega_{\alpha}$ characterizing the strength of the HDSC background, namely,  
$\Omega_\alpha =0,0.1,0.2$. 
The upper left graph of Fig. \ref{Figure678} (corresponds to $\Omega_k > 0$) and the upper right graph of Fig. \ref{Figure678} ($\Omega_k = 0$) 
show the emergence of the decelerating phase in the early universe before the accelerating phase ensues for all three values of $\Omega_{\alpha}$. However, from the lower graph of Fig.  \ref{Figure678}  (corresponds to $\Omega_k<0$), we find that the condensate can ameliorate the singularity in $q(a)$ in the conventional case without the $\alpha R^2$ term. This is clear from eqn. (\ref{q}) where with $\Omega_\alpha=0$, $q$ will become singular at $a/a_0={\sqrt{\Omega_k/\Omega_\Lambda }}$ for negative $\Omega_k$.

Let us recall that in conventional cosmology, a decelerating phase can appear only if matter is introduced from outside. However, in our case, no external matter is introduced and this decelerating phase is generated solely by the HDSC (coming from the $\alpha R^2$ term). We speculate that the HDSC might be identified as a new kind of matter candidate having radiation-like behaviour. The upper left and right plots of Fig. \ref{Figure678} show how the strength of the TC condensate affects the behavior of $q$ through different curves that merge with the $\Omega_\alpha =0$ curve asymptotically. 
However, the situation is very different for $\Omega_k=-0.1$ as depicted in  the lower graph of Fig. \ref{Figure678} . In small $a$ sector, there appears a discontinuity in $q$ for $\Omega_\alpha =0$ (standard GR, no $R^2$-term hence no condensate) that is smoothed out for non-zero $\Omega_\alpha $ ($R^2$-term with condensate effect) and finally all curves asymptotically saturate to a negative $q$ at large $a$.

\section{Summary and Conclusions}
\label{sec-conclusion}

The main novel idea presented in this work is that, the additional higher derivative scalar excitation (besides the conventional spin-$2$ graviton), generated by the $R^2$ correction to Einstein gravity, can lead to a lowest energy condensate-like state (HDSC) with an explicit periodic time dependence. As mentioned in this article, this HDSC is clearly reminiscent of the newly proposed idea of Time Crystal-like (TC) state  \cite{Wilczek:2012jt} (also see \cite{Shapere:2012nq,Ghosh:2012yj,Sacha:2017fqe}). As argued by several investigators  TC could have some effects on the cosmological dynamics \cite{Bains:2015gpv,Vacaru:2018mmd,Addazi:2018fyo,Feng:2018qnx,Das:2018ylw,Yoshida:2019pgn,Li:2019laq,Li:2019gfi,Bubuianu:2019gbo}. However, an important qualitative distinction between TC as introduced in cosmology and the  HDSC, induced by quadratic $R^2$ term, as considered here, needs to be emphasized. Recall that in the previous works, the TC feature was incorporated in the externally introduced matter whereas in our framework the TC condensate is generated within the $R^2$-gravity sector.\footnote{Let us note that  the present work does not include an externally introduced matter sector, other than the cosmological constant. However,  since  current observational data are in favor of the $\Lambda$CDM cosmological model, it will be interesting to  include additional  matter sector in eqn. (\ref{4}). Obviously, the TC condensate background will remain but the matter degrees of freedom can have non-trivial effect on the TC parameters and subsequently affect the cosmological evolution.} In the present case, the HDSC is generated internally from the combination of the metric tensor components \textit{iff} $R^2$-term in the gravity action exists.  

{As has been mentioned throughout, we exploit the property of $f (R) \sim R + \alpha R^2$ that it gives rise to a decoupled system of conventional gravity and a higher derivative scalar sector and propose a model where  the former evolves in a background of the latter, which enjoys a HDSC phase. Thus, the energy-momentum tensor for the corresponding HDSC acts as a source for cosmological dynamics,  characterized by the FLRW line element.  We found that the scale factor of the FLRW universe can be analytically solved under some approximations and hence other cosmological parameters. The behaviour of the cosmological parameters, namely, the Hubble rate ($H (a)$) and the deceleration parameter ($q (a)$) have been graphically presented (see Figs. \ref{Figure1}, \ref{Figure234}, \ref{Figure5}, \ref{Figure678}) for  different spatial geometries of the universe and for different strengths of the HDSC through $\alpha$ generated by the $\alpha R^2$-term in the gravitational action. We found some interesting observations that clearly report that the HDSC can significantly affect the cosmological dynamics. } 

{From the evolution of the Hubble rate  for a closed universe (see the lower graph of Fig. \ref{Figure234}),  we notice that the non-zero values of $\Omega_\alpha$ generate completely different profiles compared to $\Omega_\alpha = 0$ (no HDSC),  the former ones have a minimum before saturating for large $a$.  On the other hand, from Fig.  \ref{Figure5}, we notice that irrespective of the spatial geometry of the universe, the HDSC generates a decelerating phase before the acceleration of the universe starts. This indicates that after  the early decelerating phase of the universe, it enters into the late-time accelerating phase. As the non-trivial $\Omega_{\alpha}$ offers some interesting results independent of the curvature of the universe, therefore, in Fig. \ref{Figure678} we further investigated how the evolution of the deceleration parameter depends on  various strengths of the HDSC quantified through $\Omega_{\alpha}$ for different spatial geometries.  For $\Omega_k > 0$ (upper left plot of Fig. \ref{Figure678}), $\Omega_k = 0$ (upper right plot of Fig. \ref{Figure678}) universe enters into the early accelerating phase before a decelerating phase and this remains true for different 
values of $\Omega_{\alpha}$. However, for the closed universe (corresponds to the lower graph of Fig.  \ref{Figure678}), we find that the HDSC can avoid the singularity in $q(a)$ that appears  in the conventional case without 
the $\alpha R^2$ term.} 

The final take home messages of our analysis are the following: 

\begin{itemize}

    \item {The Higher Derivative Scalar Condensate, considered here,  appears purely as a geometric effect. In this sense this time dependent condensate, with the inherent time scale, is a feature of the spacetime itself. This hints at the possibility of the time translation invariant spacetime   getting reduced to one with a time periodicity.  }
    
    \item {The generic ghost problem in higher order gravity theories is absent in $R^2$-gravity which, however, is still plagued with the (relatively harmless) additional ({\it{spurious}}) scalar degree of freedom. In the $R^2$-gravity framework, as we have shown, this extra scalar can act as a condensate, that replaces the vacuum, forms a stable background for conventional gravity, leading to possible improvements with explicit predictions. }

    \item  A radiation-like feature (see eqn. (\ref{hubble-explicit})) can be obtained without any matter fields. This is a very interesting outcome since purely from the geometric corrections, one can realize a radiation-like fluid.

\end{itemize}

Thus, following the existing results and the present outcomes in the context of late and early universe, we anticipate that the physics of HDSC needs considerable attention in the cosmological dynamics. In particular, the existence of some radiation-like fluid extracted (purely) out of the geometrical sector strongly highlights this fact. The appearance of this extra relativistic species due to the geometric corrections may indicate the presence of sterile neutrinos in the universe sector \cite{Dodelson:1993je,Abazajian:2012ys,Anchordoqui:2012qu,Jacques:2013xr,DiValentino:2013qma}, and this might have some appealing consequences in the context of  Hubble tension. It has been observed that the presence of sterile neutrinos could alleviate the Hubble constant tension \cite{Carneiro:2018xwq,Gelmini:2019deq,Gelmini:2020ekg,DiValentino:2021rjj} (see also section 7 of \cite{DiValentino:2021izs} for  extensive details).  Additionally, one may naturally wonder whether HDSC may lead to some geometrical dark energy in the early universe (early dark energy fluid)~\cite{Poulin:2018cxd}, or it could play an active role in explaining the cosmological tensions \cite{DiValentino:2021izs,Perivolaropoulos:2021jda,Kamionkowski:2022pkx}. Moreover, as the present cosmological framework is purely built from the geometric sector, therefore, one can further investigate whether the finite time future singularities appearing in various cosmological theories can be avoided in this new geometrical context \cite{Caldwell:2003vq,Nojiri:2005sr,Nojiri:2005sx}.

Furthermore, we would like to mention a recent work \cite{Duary:2023nnf} showing a connection between the sign flip of the deceleration parameter and a phase transition. It might be worthwhile to look for similar correlation in the present context.

Finally, we recall that the HDSC considered here originated from the linearised version of the Einstein gravity together with the $R^2$-term. How the explicit nature of this condensate structure will be affected by non-linear effects is indeed an open problem, although we believe that the generic features of the HDSC considered here can be quantitatively affected by non-linearity, but not in a qualitative way. There is no doubt that being an  emerging field, understanding the nature and the effects of HDSC,  could open new windows in cosmology and astrophysics. It will be interesting to explore further the effects of HDSC in alternative gravitational theories other than the quadratic $f(R)$ gravity.  We hope to investigate some of them in near future.

\section{Acknowledgments}
The authors thank the anonymous referee for the constructive comments that helped us to improve the quality
of the manuscript. 
RKD  acknowledges Naresh Saha and Joydeep Majhi for helpful discussions. SP acknowledges the financial support from the  Department of Science and Technology (DST), Govt. of India under the Scheme   ``Fund for Improvement of S\&T Infrastructure (FIST)'' (File No. SR/FST/MS-I/2019/41).

\bibliography{biblio.bib}

\end{document}